\documentclass[showpacs,aps,prb,floats,twocolumn]{revtex4}

\usepackage{graphicx}
\usepackage{braket}
\usepackage{color}
\usepackage{amsmath}
\usepackage{amssymb}
\usepackage{hyperref}

\begin{document}

\newcommand{\sigp}{\sigma^+}
\newcommand{\sigm}{\sigma^-}
\newcommand{\Gspph}{\Gamma^{\sigma^+}_0}
\newcommand{\Gsmph}{\Gamma^{\sigma^-}_0}
\newcommand{\Gcdph}{\Gamma^\mathrm{cd}_0}
\newcommand{\vc}[1]{{\boldsymbol{\mathrm{#1}}}}
\newcommand{\comment}[1]{}
\newcommand{\remove}[1]{}
\newcommand{\quot}[1]{\textquotedblleft{}#1\textquotedblright}
\newcommand{\un}{\mathrm}

\newcommand{\blue}[1]{{\color{blue}#1}}
\newcommand{\red}[1]{{\color{red}#1}}

\title{Polaron master equation theory of pulse driven phonon-assisted population inversion and single photon emission from  quantum dot excitons}
\author{Ross Manson}
\email{ross.manson@queensu.ca}
\author{Kaushik Roy-Choudhury}
\author{Stephen Hughes}
\affiliation{\hspace{-40pt}Department of Physics, Engineering Physics, and Astronomy, Queen's University, Kingston, Ontario K7L 3N6, Canada\hspace{-40pt}}
\date{\today}

\begin{abstract}
We introduce an intuitive and semi-analytical polaron master equation approach to model pulse-driven population inversion and emitted single photons from  a  quantum dot exciton.
The master equation theory allows one to identify  important phonon-induced scattering rates analytically, and fully includes the role of the time-dependent pump field.
As an application of the theory, we first study a quantum dot driven by a time-varying laser pulse on and off resonance, showing the population inversion caused by acoustic phonon emission in  direct agreement with recent experiment of Quilter {\em et al.}, Phys Rev Lett {\bf 114}, 137401 (2015). We then model quantum dots in weakly coupled cavities and show the difference in population response between exciton-driven and cavity-driven systems.
Finally, we  assess the nonresonant phonon-assisted loading scheme with a quantum dot resonantly coupled to a cavity as a deterministic single photon source, and compare and contrast the important figures of merit with direct Rabi oscillation of the population using a resonant $\pi$ pulse, where the latter is shown to be much more efficient.
\end{abstract}
\maketitle

\section{Introduction}
Optically-driven semiconductor quantum dots (QDs) provide an exciting platform for research with a variety of useful properties in optoelectronics~\cite{PhysRevB.78.153309, PhysRevB.86.241304}. The spatial confinement in three dimensions of electron\textendash{}hole pairs (excitons) creates a system with discrete energy levels, similar to an atom but with practical advantages including longer stability and tunable radiative decay times. When the lowest two energy levels are isolated from the others, the result is a persistent solid-state two-level system or quantum bit, potentially usable in quantum information processing protocols~\cite{Knill2001,Bose2014} and for on-chip  nanophotonics~\cite{RevModPhys.87.347}.

Driving a QD with a resonantly-tuned optical field creates and destroys an electron\textendash{}hole pair coherently, leading to laser-driven excitonic Rabi oscillations at a frequency proportional to the pump strength of the laser. The dynamics of a QD in a semiconductor crystal, including photonic crystal 
cavities~\cite{PhysRevB.82.045308} and micropillar cavities~\cite{PhysRevLett.106.247402},  are complicated by the presence of phonons~\cite{PhysRevB.65.235311}, usually causing unwanted decoherence. In particular, longitudinal acoustic (LA) phonons~\cite{PhysRevLett.104.017402} cause the Rabi oscillations to decay over time, especially with strong driving fields.

Achieving controlled population inversion through coherent excitation is a challenge, but useful for applications such as single-photon sources~\cite{PhysRevB.82.045308}. In the realm of pulsed excitation, inversion can be achieved by selecting the pulse strength and duration to produce an odd number of Rabi half-oscillations, leaving the QD in its excited state (i.e., with a population above 0.5)~\cite{1367-2630-15-5-053039} if it began in the ground state. However, phonon-induced dephasing and damping makes coherent control difficult, and this scheme is also very sensitive to the laser detuning and the pump strength.

Recently, another method of inversion has been proposed~\cite{PhysRevLett.110.147401} and implemented~\cite{PhysRevLett.114.137401, PhysRevB.90.241404, PhysRevB.91.161302} which actually benefits from phonon coupling. In this case the dot is driven slightly above resonance (by about $1~\un{meV}$) and excited incoherently from the phonon bath through phonon emission. This method of inversion is likely more robust with respect to variations in the pump strength and laser detuning, which makes it a promising method for  applications.
Another benefit for single-photon applications is that this scheme allows one to  spectrally separate the pump field from the single photons emitted by the excited exciton. Phonon-assisted incoherent excitation is also possible with continuous-wave (cw) driving~\cite{1367-2630-15-5-053039,Physics.8.29}, and through coupling to an off-resonant quantized cavity mode \cite{PhysRevLett.107.193601}.

Quantum dot excitons driven to  inversion are of particular interest as single photon sources~\cite{1367-2630-6-1-089, 0034-4885-75-12-126503, Shields2007, Michler22122000, PhysRevLett.89.233602, PhysRevX.2.011014, doi:10.1021/nl503081n, :/content/aip/journal/apl/96/1/10.1063/1.3284514, He2013}. In addition, coupling from the QD to an integrated optical cavity allows the relaxation of the QD from the excited state to the ground state to produce a single photon that can be collected and emitted more efficiently through the Purcell effect. Some key attributes that make a single photon source useful are the indistinguishability of the photons produced and the collection efficiency~\cite{PhysRevA.69.032305}; an ideal source would produce an indistinguishable single photon, on demand, each time an optical pulse is sent in. Realistic single photon sources strive, then, to have a high probability of producing a single photon, a narrow time window during which the photon might be produced, and a high indistinguishability between each photon and the subsequent ones~\cite{PhysRevA.69.032305}, but this is an ongoing challenge. Very recently 
Somaschi {\em et al}.~\cite{Somaschi2015} has used
resonant $\pi$ pulse excitation of QDs in  micropillar cavities and demonstrated near-optical single photon sources, where the pump field is spectrally filtered from the emitted single photons. 

In this work, we introduce and apply an efficient master equation (ME) model to   examine a QD driven by a time-varying laser pulse on and off resonance, showing the population inversion caused by acoustic phonon emission~\cite{1367-2630-15-5-053039,PhysRevLett.110.147401,PhysRevLett.114.137401}. Our framework uses and exploits a polaron ME approach~\cite{McCutcheon2010, PhysRevB.85.115309}, which accurately determines the nature of the phonon coupling using a quasiparticle (polaron) transform. The full polaron ME~\cite{McCutcheon2010, PhysRevB.85.115309} is restructured analytically here to express phonon-mediated scattering in terms of distinct phonon rates corresponding to different physical processes. This allows a much faster computation and deeper understanding of the underlying mechanisms, which complex numerical approaches, such as path integral techniques~\cite{PhysRevLett.110.147401} and correlation expansion approaches~\cite{PhysRevLett.91.127401}, can sometimes obscure. The polaron ME is also simpler to use than than   variational ME approaches~\cite{PhysRevB.84.081305}, which are valid for a wider range of pump strengths than either the polaron~\cite{McCutcheon2010} or weak-coupling~\cite{PhysRevLett.108.017401} approaches; however, the polaron theory is valid for all parameters considered here and is more suitable to use with general pulse shapes. Master equation techniques also allow one to easily calculate experimentally relevant quantities in quantum optics, such as the two-time correlation functions\cite{PhysRevA.69.032305} and the emitted resonance fluorescence spectrum~\cite{PhysRevLett.110.217401,Carmichael1}. Moreover, this theoretical approach fully incorporates the influence of the optical field on the phonon rates, making it valid in a higher-pump regime, in contrast to a previous effective phonon ME approach~\cite{Ulhaq2013}, to which it reduces in the low-pump limit. Using this semi-analytical ME theory, we first directly reproduce recent experimental trends for pulse-excited population inversion of single QDs~\cite{PhysRevLett.114.137401} and explain the underlying physics in terms of the phonon-assisted scattering rates.
We then use the same technique to study the population inversion of a QD under cavity driving and coupling, using a cavity that is weakly coupled to the dot and tuned to a higher frequency than the exciton resonance. Finally, we investigate the feasibility of using the regime of phonon-assisted incoherent excitation as a deterministic single photon source, assessing the important figures of merit~\cite{PhysRevA.69.032305}. We also provide an Appendix,  where we present the derivations of the analytical scattering rates for the phonon-mediated scattering processes.

\section{THEORETICAL formalism }
\subsection{Hamiltonian for the quantum dot exciton with phonon coupling}

The QD is modelled as a two-level system with ground state $\ket{g}$, excited state $\ket{e}$, and raising and lowering operators $\sigp = \ket{e}\bra{g}$ and $\sigm = \ket{g}\bra{e}$. We assume the QD is driven by an optical pulse with Rabi frequency $\Omega(t) = \Omega_p e^{-t^2/\tau_p^2}$, where $\Omega_p$ is the peak Rabi frequency  and $\tau_p$ is a characteristic pulse halfwidth. The $1/e$ pulse fullwidth has a value of $2\tau_p$. Another useful metric for describing light-matter interactions with pulses is the pulse area, defined as $\Theta = \int_{-\infty}^{\infty}\Omega(t)dt$, where $\Theta = \sqrt{\pi}\tau_p\Omega_p$ for a Gaussian pulse.

The QD exciton transition is coupled to a bath of acoustic phonons, which is modelled as a collection of harmonic oscillators with wave vector $\vc{q}$, creation and annihilation operators $b^\dagger_\vc{q}$ and $b_\vc{q}$, and frequency $\omega_\vc{q} = c_s|\vc{q}|$, where $c_s$ is the speed of sound. Each phonon mode couples to the exciton with strength $\lambda_\vc{q}$ (assumed real).
In an interaction picture, rotating at the laser frequency $\omega_L$, we express the total Hamiltonian as
\begin{equation}
\begin{split}
H(t) =\:&\frac{1}{2}\hbar\Omega(t)\left(\sigp + \sigm\right) - \hbar\Delta_{Lx}\sigp\sigm \\ &+ \sum_{\vc{q}} \hbar\omega_{\vc{q}} b^\dagger_{\vc{q}} b_{\vc{q}} + \sigp\sigm\sum_{\vc{q}}\hbar\lambda_{\vc{q}}\left(b_{\vc{q}}+ b^\dagger_{\vc{q}}\right),
\end{split}
\end{equation}
where $\Delta_{Lx} = \omega_L - \omega_x$ is the detuning between the laser frequency and exciton frequency. 

Next, a transform is made to the polaron frame~\cite{PhysRevB.85.115309,McCutcheon2010}, using the operator $P = \sigp\sigm\sum_{\vc{q}}\frac{\lambda_\vc{q}}{\omega_\vc{q}}(b_{\vc{q}}- b^\dagger_{\vc{q}})$, to produce the polaron-transformed Hamiltonian $H^\prime = e^PHe^{-P}$. This rearranges the Hamiltonian in terms of polarons, which are a hybrid excitation of excitons and phonons, rather than one or the other alone. This allows a better separation into system parts and bath parts, while fully recovering the solution of the independent boson model~\cite{PhysRevB.65.235311,PhysRevB.85.115309,Hohenester2006}.
This transformed Hamiltonian can then be split into system, bath, and interaction parts, $H^\prime = H_S^\prime + H_B^\prime + H_I^\prime$, where~\cite{PhysRevB.85.115309, McCutcheon2010}
\begin{align}
H_S^\prime &= \hbar(-\Delta_{Lx} - \Delta_P)\sigp\sigm + \langle B\rangle X_g(t), \\
H_B^\prime &= \sum_\vc{q} \hbar\omega_\vc{q} b^\dagger_{\vc{q}} b_{\vc{q}}, \\
H_I^\prime &= X_g(t)\zeta_g + X_u(t)\zeta_u.
\end{align}
Here the phonon-modified system operators are given by
$
X_g(t) = \frac{1}{2}\hbar\Omega(t)(\sigp+\sigm) $ and $
X_u(t) = \frac{1}{2}i\hbar\Omega(t)(\sigp-\sigm),
$
and the bath-induced fluctuation operators are
$
\zeta_g = \frac{1}{2}(B_++B_--2\braket{B}) $ and $
\zeta_u = \frac{1}{2i}(B_+-B_-).
$
The coherent-state phonon displacement operators are
$
B_\pm = \exp\left(\pm\sum_\vc{q}\frac{\lambda_q}{\omega_q}(b_\vc{q}-b_\vc{q}^\dagger)\right),
$
with expectation value $\braket{B_+} = \braket{B_-} \equiv \braket{B}$. Below we will absorb the polaron shift $\Delta_P = \sum_{\vc{q}}\lambda_\vc{q}^2/\omega_\vc{q}$ into the definition of $\omega_x$ and thus it can be effectively ignored hereafter.

\subsection{Full polaron master equation with time-dependent pump fields}

{Since the exciton system is coupled to a phonon bath, it is beneficial to use a ME to solve the dynamics, treating the QD as a two-level excitonic system and the phonons as a perturbation. A weak coupling approach is often used to describe the phonons~\cite{PhysRevLett.108.017401}; however, in some regimes such as strong exciton\textendash{}phonon coupling and higher temperatures, it is not sufficient to describe the phonon-mediated dynamics of the system~\cite{McCutcheon2010,PhysRevB.84.081305}. Instead, the polaron transform is used, creating a hybrid excitation (polarons) which incorporates multi-phonon effects.} 
One advantage of this approach is that it fully recovers the independent boson model in the appropriate limit
\cite{PhysRevB.65.195313, PhysRevLett.91.127401}, a well-known solution for linear excitation. 
With nonlinear excitation, the Born\textendash{}Markov approximation can be performed with respect to the polaron-transformed perturbation.
The polaron-transformed Hamiltonian is used to produce a polaron ME\cite{PhysRevB.85.115309, McCutcheon2010}, describing the dynamics of the system density operator $\rho(t)\remove{ = \sum_i p_i \ket{\psi_i}\bra{\psi_i}}$, in the presence of the phonon bath with density operator $\rho_B$. This is presented here in the form obtained after a Born\textendash{}Markov approximation (time-convolutionless or time-local form), with an additional Markov approximation made in the phonon relaxation time~\cite{PhysRevB.85.115309}. 
In a ME approach, we can also easily include important background processes, such as radiative decay $\gamma$ and pure dephasing $\gamma^\prime$, as Lindblad rates corresponding to ME terms defined as $\mathcal{L}[\hat{O}]\rho = 2\hat{O}\rho\hat{O}^\dagger - \hat{O}^\dagger\hat{O}\rho - \rho\hat{O}^\dagger\hat{O}$, where
$\hat O$ is the relevant operator. 
Including such additional zero phonon line processes,
we obtain the full polaron ME of interest
\begin{equation}
\label{FPME}
\frac{\partial\rho}{\partial t} = \frac{1}{i\hbar}[H_S^\prime(t), \rho(t)] + \frac{\gamma}{2}\mathcal{L}[\sigm]\rho + \frac{\gamma^\prime}{2}\mathcal{L}[\sigp\sigm]\rho + \mathcal{L}_\mathrm{ph}\rho,
\end{equation}
with
\begin{align}
\mathcal{L}_\mathrm{ph}\rho = &-\frac{1}{\hbar^2}\int_0^\infty\sum_{m=g,u}d\tau \\
&\times \left(G_m(\tau)[X_m(t), X_m(t,\tau)\rho(t)] + \operatorname{H.c.}\right), \nonumber
\end{align}
where $X_m(t,\tau) = e^{-iH_S^\prime(t)\tau/\hbar}X_m(t)e^{iH_S^\prime(t)\tau/\hbar}$, and the polaron Green functions are given by~\cite{PhysRevB.85.115309}
$G_g(\tau) = \braket{B}^2\left(\cosh(\phi(\tau))-1\right)$ and 
$G_u(\tau) = \braket{B}^2\sinh(\phi(\tau))$.

The phonon-induced dephasing that distinguishes optically driven QD systems from isolated atomic systems is primarily due to LA phonons through a deformation potential~\cite{PhysRevLett.104.017402, PhysRevX.1.021009, PhysRevB.65.195313}. The phonons are described by phonon correlation and spectral functions, which are evaluated using a continuum approximation that assumes the phonon modes are close together compared to the other frequency differences. In that case the phonon spectral function can be written as 
$
J(\omega) = \alpha_P \omega^3 \exp\left(-\omega^2/2\omega_b^2\right),
$
and the phonon correlation function is obtained from
\begin{equation}
\phi(\tau) = \int_{0}^{\infty} \frac{J(\omega)}{\omega^2} 
\left[\coth\left(\frac{\hbar\omega}{2k_B T}\right) \cos(\omega\tau)-i\sin(\omega\tau)\right] d\omega,
\end{equation}
while the bath operator expectation value is given by $\braket{B} = \exp\left[-\frac{1}{2} \int_0^\infty \frac{J(\omega)}{\omega^2} \coth(\hbar\omega/2k_BT)\right] = \exp(-\phi(0)/2)$.
The parameters $\alpha_P$  and $\omega_b$  are the electron\textendash{}phonon coupling strength and the phonon cutoff frequency, respectively. These are material properties of the QD and phonon environment, and their values are typically determined by fitting to experiments on single QDs\cite{PhysRevLett.104.017402, PhysRevB.83.165313}.

\subsection{Simplification of the polaron master equation with analytical scattering rates}

The full polaron ME (Eq.~\ref{FPME}) contains, in its commutators, various permutations of $\sigp$, $\sigm$, and $\rho$. Thus it can be expanded to produce a set of phonon-mediated processes with well-defined rates, each multiplied by one such permutation. This results in a more transparent analytical form of the ME, which can be solved numerically much more quickly than the full form which involves complex operator exponentials at two times. The analytical form also allows considerable insight into the different processes, such as phonon-mediated photon emission and absorption. No approximations beyond those used to derive Eq.~\ref{FPME} need to be made to simplify the ME, but non-contributing terms can be omitted at the end.

The commutators in the final term of Eq.~\ref{FPME}, $\mathcal{L}_\mathrm{ph}\rho$, are expanded and rearranged using fermion operator identities, so that we can rewrite the ME as (see Appendix)
\begin{align}
\label{APME}
\frac{\partial\rho(t)}{\partial t} =
 &-\frac{i}{\hbar}\left[H_S^\prime,\rho \right]
  + \frac{\gamma}{2}\mathcal{L}[\sigm]\rho
  + \frac{\gamma^\prime}{2}\mathcal{L}[\sigp\sigm]\rho \nonumber \\
 &+ \frac{\Gamma^{\sigma^+}}{2}\mathcal{L}[\sigp]\rho
  - \Gamma^\mathrm{cd}(\sigp\rho\sigp + \sigm\rho\sigm) \nonumber \\
 &+ \frac{\Gamma^{\sigma^-}}{2}\mathcal{L}[\sigm]\rho
  -i\Gamma^\mathrm{sd}(\sigp\rho\sigp - \sigm\rho\sigm) \nonumber \\
 &-\left(i \Gamma_u(\sigp\sigm\rho\sigp + \sigm\rho - \sigp\sigm\rho\sigm) + \operatorname{H.c.}\right) \nonumber \\
 &-\left(\Gamma_g(\sigp\sigm\rho\sigp - \sigm\rho + \sigp\sigm\rho\sigm) + \operatorname{H.c.}\right) \nonumber \\
 &+ i\Delta^{\sigp\sigm} [\sigp\sigm,\rho].
\end{align}
%
%
Of the seven phonon rates, four are significant in the parameter regimes considered here, and are given by
\begin{align}
\Gamma^{\sigma^+ / \sigma^-}(t) &= \frac{\Omega_R(t)^2}{2}\int_{0}^{\infty}\Bigg(\operatorname{Re}\bigg\{(\cosh(\phi(\tau))-1) f(t,\tau) \nonumber \\
&+ \sinh(\phi(\tau))\cos(\eta(t)\tau)\bigg\} \label{Gsp_t} \nonumber \\
&\mp \operatorname{Im}\left\{(e^{\phi(\tau)}-1)
\frac{\Delta_{Lx}\sin(\eta(t)\tau)}{\eta(t)}\right\}\Bigg)\,d\tau,
 \\
\Gamma^\mathrm{cd}(t) &= \frac{\Omega_R(t)^2}{2}\int_{0}^{\infty} \operatorname{Re}\bigg\{\sinh(\phi(\tau))\cos(\eta(t)\tau) \nonumber \\
&- (\cosh(\phi(\tau))-1) f(t,\tau) \bigg\} d\tau, \label{Gcd_t} \\
\Gamma_u(t) &= \frac{\Omega_R(t)^3}{2\eta(t)}\int_{0}^{\infty}\sinh(\phi(\tau))\sin(\eta(t)\tau)d\tau, \label{Gu_t}
\end{align}
where $f(t,\tau) = \frac{\Delta_{Lx}^2\cos(\eta(t)\tau)+\Omega_R(t)^2}{\eta(t)^2}$, and $\eta(t) = \sqrt{\Omega_R(t)^2 + \Delta_{Lx}^2}$ with the polaron-shifted Rabi frequency:  $\Omega_R(t) = \braket{B}\Omega(t)$.
The other three rates are found to be negligible for our studies below, but are given by
$\Gamma_g(t) = \frac{\Omega_R(t)^3\Delta_{Lx}}{2\eta(t)^2}\int_0^\infty(\cosh(\phi(\tau))-1)(1-\cos(\eta(t)\tau)) d\tau,$
$\Gamma^\mathrm{sd}(t) = \frac{\Omega_R(t)^2}{2}\int_0^\infty\operatorname{Re}\left\{1-e^{-\phi(\tau)}\right\}\frac{\Delta_{Lx}\sin(\eta(t)\tau)}{\eta(t)}d\tau,$
and $\Delta^{\sigp\sigm}(t) = \frac{\Omega_R(t)^2\Delta_{Lx}}{2\eta(t)}\int_0^\infty\operatorname{Re}\left\{e^{\phi(\tau)}-1\right\}\sin(\eta(t)\tau)d\tau.$

In the case of a weak optical drive $\Omega_R(t) \ll \Delta_{Lx}$, Eq.~\ref{APME} reduces to the effective phonon ME~\cite{Ulhaq2013, PhysRevX.1.021009} used before for weak cw drives, given by
\begin{align}
\label{EPME}
\frac{\partial\rho(t)}{\partial t}& = -\frac{i}{\hbar}\left[\hat{H}_S,\rho \right] + \frac{\gamma}{2}\mathcal{L}[\sigm]\rho + \frac{\gamma^\prime}{2}\mathcal{L}[\sigp\sigm]\rho \\
&\hspace{-20pt}+ \frac{\Gspph}{2}\mathcal{L}[\sigp]\rho + \frac{\Gsmph}{2}\mathcal{L}[\sigm]\rho - \Gcdph(\sigp\rho\sigp + \sigm\rho\sigm) \nonumber,
\end{align}
where the phonon rates that survive take the somewhat simpler form
\begin{align}
\Gamma^{\sigma^+ / \sigma^-}_\mathrm{0}(t)\! =\! \frac{(\Omega_R(t))^2}{2}\!\int_{0}^{\infty}\operatorname{Re}\left\{ e^{\pm i\Delta_{Lx}\tau}\left(e^{\phi(\tau)}-1\right) \right\}d\tau, \label{Gspph_t} \\
\Gcdph(t)\! =\! \frac{(\Omega_R(t))^2}{2}\!\int_{0}^{\infty}\operatorname{Re}\left\{ \cos(\Delta_{Lx}\tau)\left(1-e^{-\phi(\tau)}\right) \right\}d\tau. \label{Gcdph_t}
\end{align}
This effective ME applies when the optical drive is weak, particularly in the cw case when the Rabi frequency need not exceed $\sim0.1~\un{meV}$ to cause phonon-assisted population inversion~\cite{1367-2630-15-5-053039}; it is also  valid in the pulse-driven case with appropriate experimental parameters~\cite{PhysRevLett.114.137401}, such as $2\tau_p = 20.2~\un{ps}$ and $\Theta = 7.24\pi$, where $\Omega_p = 0.84~\un{meV}$. Equation~\ref{EPME} has been used to successfully explain experiments in resonance fluorescence~\cite{Ulhaq2013} as well as phonon-assisted exciton loading~\cite{PhysRevB.86.241304}.

However, this form does not hold in the regime where short pulses are strong enough to create exciton inversion ($\Omega_p \gtrsim 1~\un{meV}$). In the high pump power regime Eq.~\ref{APME} is used instead, though we 
will also compare with Eq.~\ref{EPME}.

\subsection{Numerical approach to solving the pulse excited polaron master equation}

The polaron ME (Eq.~\ref{APME}) is solved using a numerical ODE solving method to produce the system density operator $\rho(t)$, given an initial condition $\rho(0)$ corresponding to the QD in its ground state\comment{and the phonon bath at temperature $T$}. This time-dependent state is then used to calculate expectation values of operators, such as the excited-state population given by $N_x(t) = \braket{\sigp\sigm}(t) = \operatorname{Tr}\left\{\rho(t)\sigp\sigm\right\}$. This can be used, e.g., to assess whether a state population has reached inversion ($N_x > 0.5$).

\section{Results for pulse excited population inversion}

\subsection{Pulse excited phonon-mediated scattering rates}
\label{results_rates}

We choose parameters common in InGaAs/GaAs QDs and similar to those used in Ref.~\onlinecite{PhysRevLett.114.137401}. The pulsewidth is $2\tau_p = 20.2~\un{ps}$ (16.8 ps FWHM), the phonon bath temperature is $T = 4.2~\un{K}$, and we take $\alpha_p = 0.03~\un{ps^2}$, $\omega_b = 1~\un{meV}$, $\gamma = 2~\un{\mu eV}$, $\gamma^\prime = 2~\un{\mu eV}$. The laser--exciton detuning is often varied, but when it is held fixed, we choose $\Delta_{Lx} = +0.83~\un{meV}$, a near-optimal detuning for phonon-assisted inversion~\cite{1367-2630-15-5-053039} and the same detuning studied in Ref.~\onlinecite{PhysRevLett.114.137401}.

\begin{figure}[t]
\includegraphics[width=\linewidth]{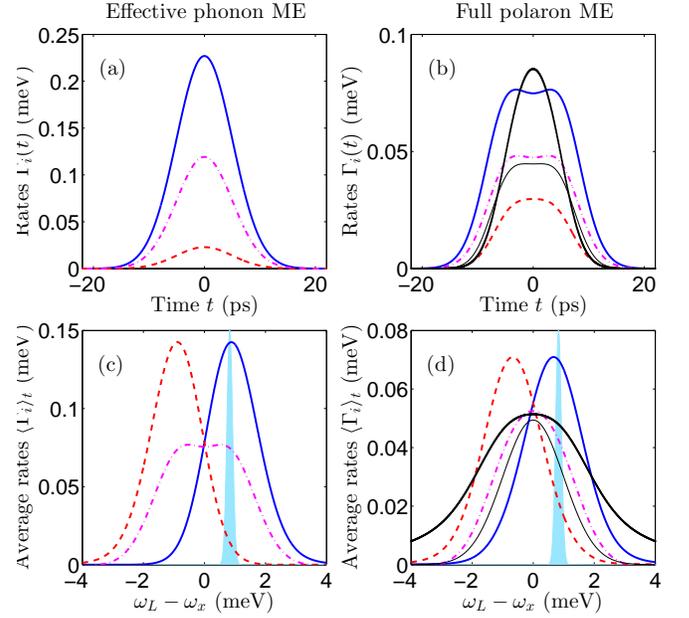}
\caption{\footnotesize{(Color online). Phonon rates in two models (see text), as a function of time and detuning, for $\Theta = 16\pi$.
(a) Effective phonon ME rates (Eq.~\ref{EPME}) for $\Delta_{Lx} = +0.83~\un{meV}$: $\Gspph(t)$ (Eq.~\ref{Gspph_t}, blue solid),  $\Gsmph(t)$ (Eq.~\ref{Gspph_t}, red dashed),  $\Gcdph(t)$ (Eq.~\ref{Gcdph_t}, magenta dot-dashed). \
(b) Full polaron ME rates (Eq.~\ref{APME}) for $\Delta_{Lx} = +0.83~\un{meV}$: $\Gamma^{\sigma^+}(t)$ (Eq.~\ref{Gsp_t}, blue solid),  $\Gamma^{\sigma^-}(t)$ (Eq.~\ref{Gsp_t}, red dashed),  $\Gamma^\mathrm{cd}(t)$ (Eq.~\ref{Gcd_t}, magenta dot-dashed), $\operatorname{Re}[\Gamma_u(t)]$ (Eq.~\ref{Gu_t}, thick black solid), $\operatorname{Im}[\Gamma_u(t)]$ (Eq.~\ref{Gu_t}, thin black solid). \
(c) Effective phonon ME rates (Eq.~\ref{EPME}), time averaged over one pulse full width. Each rate is time averaged as $\braket{\Gamma_i}_t = \int_{-\infty}^{\infty} \Gamma_i(t) dt / 2\tau_p$, and they have the same color scheme as in (a). The laser pulse in the Fourier domain with $\Delta_{Lx} = +0.83~\un{meV}$ is shown in shaded light blue in normalized units. \
(d) Full polaron ME rates (Eq.~\ref{APME}), time averaged over one pulse full width, with the same color scheme as in (b).}}
\label{phrates}
\end{figure}

The phonon rates from the analytical phonon ME (Eq.~\ref{APME}) and the effective phonon ME (Eq.~\ref{EPME}) are plotted in Fig.~\ref{phrates} for a drive strength of $\Theta = 16\pi$, which corresponds to a peak Rabi frequency of $1.85~\un{meV}$ for the chosen pulsewidth. The phonon rates' time dependence due to the time-varying pulse amplitude is shown for a detuning of $\Delta_{Lx} = 0.83~\un{meV}$, and the time-averaged phonon rates are shown as a function
of detuning. The phonon emission, absorption and cross dephasing terms take a different time and detuning dependence in the full model than in the effective model, and the term $\Gamma_u$ is only present in the full model. When the drive is weaker ($\Omega_p \ll \omega_b$, corresponding to $\Theta \lesssim 8\pi$ for these parameters), the effective phonon ME is valid and the phonon rates look the same in the two models (not shown).

Two of the phonon scattering rates, $\Gamma^{\sigma^+}$ and $\Gamma^{\sigma^-}$, highlight the physical processes that affect phonon-induced population inversion~\cite{Ulhaq2013}. Phonon emission accompanies the transfer of the laser--exciton system from a higher-energy state to a lower-energy state, while phonon absorption accompanies the transfer from a lower-energy state to a higher-energy state. At low temperatures, due to the scarcity of phonons in the bath, phonon emission is more likely than phonon absorption. Therefore, for positive detuning ($\Delta_{Lx} > 0$), the preferred process is a transfer of energy from the laser field to the exciton state along with emission of a phonon, amounting to an incoherent excitation at a rate of $\Gamma^{\sigma^+}$, rather than a transfer of energy from the exciton state to the laser field along with absorption of a phonon, a radiative decay process at a rate of $\Gamma^{\sigma^-}$. This low-temperature asymmetry is shown in Fig.~\ref{phrates}(c,d), where $\Gamma^{\sigma^+} > \Gamma^{\sigma^-}$ for a laser pulse of detuning $\Delta_{Lx} = +0.83~\un{meV}$. At higher temperatures ($k_B T \gg \hbar \omega_b$) the two phonon rates would be nearly equal, and the incoherent excitation process would not be possible, since the reverse process, radiative decay, would also take place at the same rate.

\begin{figure}[h!]
\includegraphics[width=\linewidth]{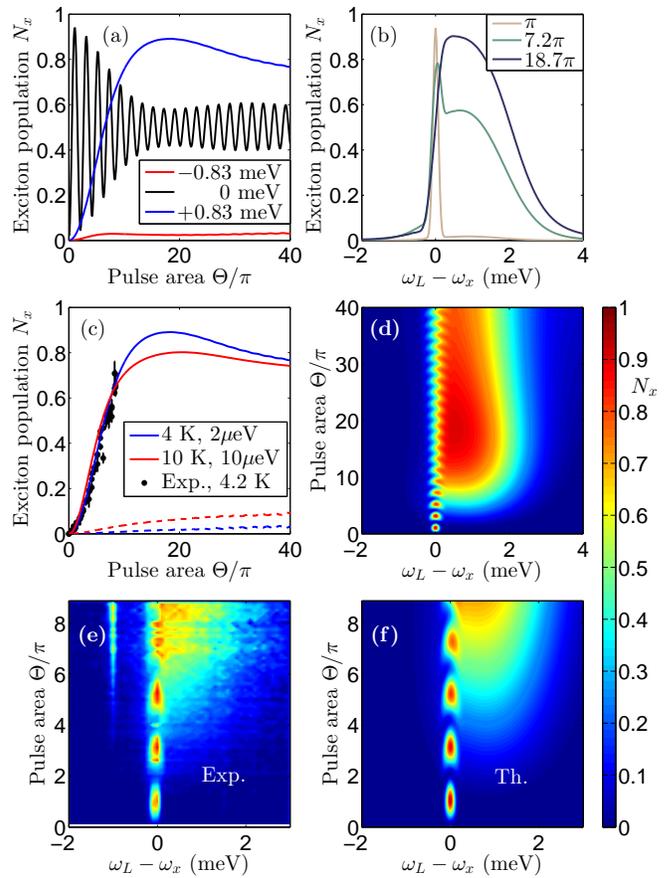}
\caption{\footnotesize{(Color online) Exciton population calculated with the analytical polaron ME. (a) Exciton population at detunings of $\Delta_{Lx} = -0.83~\un{meV}$ (red-lowest), $0~\un{meV}$ (black-middle) and $+0.83~\un{meV}$ (blue-highest), as a function of pulse area. Phonon-assisted inversion is seen when the laser is above resonance (blue). At a pulse area of $18\pi$, the population reaches a maximum of 0.9 and begins to decrease again as the very strong Rabi field (4.6 meV for $\Theta = 40\pi$) moves outside the phonon bath function. By a similar mechanism the amplitude of the Rabi rotations on resonance (black) is damped for moderate pulse areas, then increases again for very strong pulse areas. (b) Exciton population as a function of pump detuning $\hbar\Delta_{Lx}$, for pulse areas $\Theta = \pi$ (light), $7.2\pi$ (medium), $18.7\pi$ (dark). With a $\pi$-pulse, exciton inversion is seen on resonance, whereas a much stronger pulse is needed for off-resonant phonon-assisted inversion. (c) Population with $\Delta_{Lx} = 0.83~\un{meV}$, with phonon effects calculated with Eq.~\ref{APME} (solid) and phonons ignored (dashed), at a temperature and pure dephasing of $T = 4~\un{K}$ and $\gamma' = 2~\un{\mu eV}$ (blue; upper solid and lower dashed) and $T = 10~\un{K}$ and $\gamma' = 10~\un{\mu eV}$ (red; lower solid and upper dashed). {Experimental data from Ref.~\onlinecite{PhysRevLett.114.137401} (black points with error bars) agrees with the phonon theory calculated by Eq.~\ref{APME}. (d) Exciton occupation for varying detuning and pulse area, with temperature and pure dephasing of $T = 4~\un{K}$ and $\gamma' = 2~\un{\mu eV}$ as elsewhere. (e) Experimental occupation~\cite{PhysRevLett.114.137401}, with $T = 4.2~\un{K}$. (f) Theoretical occupation, as in (d) but over a smaller range of values to allow a better comparison to the experiments.}}}
\label{expth}
\end{figure}

\subsection{Pulse driven exciton populations and  connection to  the recent experiments of Quilter {\em et al}.~\cite{PhysRevLett.114.137401}}

The metric used to determine whether a QD has reached inversion is the final exciton population. In the case of a cw drive or a system without spontaneous decay, the population at $t \to \infty$ is used; however, with a pulse drive $\Omega(t)$ and a constant decay $\gamma$, the population begins to decay as soon as the pulse has ended, so the population at $t \to \infty$ is zero. Instead we take the population at a time $2\tau_p$ (one pulse fullwidth) after the center of the pulse, which avoids counting brief Rabi oscillations as inversion but effectively captures the total effect of the pulse. This population is shown in Fig.~\ref{expth} for a variety of pulse parameters. \remove{We compare our theoretical results obtained from solving Eq.~\ref{APME} to the experimental results from Quilter {\em et al.}~\cite{PhysRevLett.114.137401}.} Figure~\ref{expth}(a) shows population at three detunings and varying pulse area, while Fig.~\ref{expth}(b) holds pulse area fixed and varies detuning. Figure~\ref{expth}(c) shows the effect at different temperatures with and without phonons{, along with experimental data obtained by Quilter \emph{et al.}~\cite{PhysRevLett.114.137401}. The excellent agreement indicates that the observed off-resonant inversion is quantitatively explained by LA phonons modelled with a polaron ME formalism, even though we have made no attempt to fit the phonon spectral function for the specific QDs.} Figure~\ref{expth}(d) and (f) show the theoretical population at a range of pulse areas and detunings, and Fig.~\ref{expth}(e) shows the experimental populations from Ref.~\onlinecite{PhysRevLett.114.137401}.

\subsection{Exciton populations through cavity driving}

If the optical drive excites a coupled cavity instead of the exciton, the dynamics are somewhat different. The Hamiltonian includes the additional cavity interaction terms~\cite{1367-2630-15-5-053039}
\begin{equation}
H_c = \hbar g (\sigp a + a^\dagger\sigm) - \hbar\Delta_{Lc}a^\dagger a,
\end{equation}
where $g$ is the cavity\textendash{}dot coupling, $\Delta_{Lc} = \omega_L - \omega_c$ is the detuning between the laser drive and the cavity resonance, and $a^\dagger$ and $a$ are cavity photon creation and annihilation operators.
In addition, the cavity drive $\frac{1}{2}\hbar \Omega_c(t) (a^\dagger + a)$ replaces the exciton drive of $\frac{1}{2}\hbar \Omega(t) (\sigp + \sigm)$.

In the bad-cavity limit, the cavity can be approximated to be in a coherent state, allowing the operators $a^\dagger$ and $a$ to be replaced by $c$-numbers~\cite{1367-2630-15-5-053039}. This eliminates the cavity degrees of freedom, and one can define an effective exciton drive $\Theta_\mathrm{eff} = \int_{-\infty}^{\infty}\Omega_{\mathrm{eff}}(t)dt$ based on the cavity drive $\Theta_c = \int_{-\infty}^{\infty}\Omega_c(t)dt$. These drives are related through  
\begin{align}
\Omega_{\mathrm{eff}}(t) &= \frac{g\Omega_c(t)}{\sqrt{\kappa^2+\Delta_{Lc}^2}}, \\
\Theta_\mathrm{eff} &= \frac{g\Theta_c}{\sqrt{\kappa^2+\Delta_{Lc}^2}},
\end{align}
where $\kappa$ is the cavity decay rate, related to the cavity $Q$-factor by $\kappa = \omega_c/Q$. \comment{Three cavity drive values are chosen, $\Theta_c = 7.24\pi, 16\pi, 32\pi$.}The details of this approach are discussed in Ref.~\onlinecite{1367-2630-15-5-053039}, and we have checked numerically that the full cavity\textendash{QD} ME~\cite{PhysRevB.65.235311}, which includes the cavity operators at the system level, gives basically the same result for a weakly coupled cavity. We choose $g = 50~\un{\mu eV}$, and fix the cavity\textendash{}dot detuning at $\Delta_{cx} = 0.83~\un{meV}$. Two values of $\kappa$ are studied, $\kappa = 138~\un{\mu eV}$ and $\kappa = 620~\un{\mu eV}$, corresponding to $Q = 9000$ and $Q = 2000$ if $\omega_c = 1.24~\un{eV}$.

It is important to note that the Rabi field of the cavity drive will in general be much larger than the Rabi field of direct QD excitation, so with the laser same power,
$\Omega_c\gg \Omega$, due to resonant enhancement from the cavity mode. For example, if exciting on-resonance, then the drive intensity will be increased by about a factor of $Q$, which makes the cavity-coupled QD system advantageous for probing high field effects.

\begin{figure}[t]
\includegraphics[width=\linewidth]{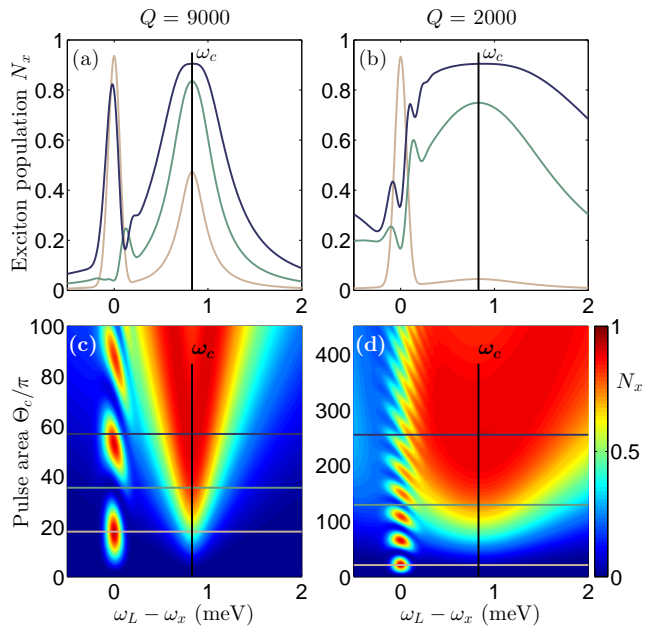}
\caption{\footnotesize{(Color online) Predicted exciton population from cavity driving, with $\Delta_{cx} = 0.83~\un{meV}$, for $\kappa = 138~\un{\mu eV}$ (left) and $\kappa = 620~\un{\mu eV}$ (right), at three selected pulse areas (top) and a wide range of pulse areas (bottom). (a) $Q = 9000$, cavity pulse areas $\Theta_c = 18\pi$ (light), $35.5\pi$ (medium) and $57\pi$ (dark). (b) $Q = 2000$, cavity pulse areas $\Theta_c = 21\pi$ (light), $129\pi$ (medium) and $255\pi$ (dark). (c) $Q = 9000$, varying pulse areas with the three selected in (a) marked. (d) $Q = 2000$, varying pulse areas with the three selected in (b) marked.}}
\label{cavitypop}
\end{figure}

The results for a cavity drive are calculated in the same way as for exciton driving, with the detuning-dependent effective exciton drive replacing the true exciton drive. The phonon rates are also calculated with similar expressions as in the exciton case (Eq.~\ref{Gsp_t}\textendash\ref{Gu_t}) but describe state transfer to and from the cavity instead of the laser field. For example, with exciton driving, the rate $\Gamma^{\sigma^+}$ describes an incoherent excitation process with state transfer from the laser to the exciton, whereas with cavity driving, the rate $\Gamma^{a\sigma^+}$ describes a state transfer from the cavity to the exciton~\cite{1367-2630-15-5-053039}.
Figure~\ref{cavitypop}(a)-(b) plot the exciton population for three different cavity drive values, while Fig.~\ref{cavitypop}(c)-(d) plot the same profile for a wide range of cavity drive strengths.
The detuning-dependent behavior of the phonon-mediated exciton population is similar to in the case of exciton driving (cf. Fig.~\ref{expth}(b)). The most noticeable difference is that the broad region of phonon-assisted inversion found above resonance in exciton driving is narrowed due to cavity filtering in the cavity-driving case and centered at $\Delta_{cx}$, with the narrowing especially pronounced for the high-$Q$ cavity. The height and number of peaks near the laser\textendash{}exciton resonance in Fig.~\ref{cavitypop}(a)-(b) depend on the particular driving strength's place in the Rabi rotations (cf. horizontal cross sections of Fig.~\ref{cavitypop}(c)-(d)). In the case of $Q = 9000$ and $\Theta_c = 57\pi$ (dark curve in Fig.~\ref{cavitypop}(a)), three peaks are visible: one from the cavity-filtered phonon-assisted inversion, and two near laser--exciton resonance from the Rabi rotations, which are asymmetric with respect to laser detuning due to the influence of the cavity on the effective drive strength. When $Q = 2000$ and $\Theta_c = 129\pi$ or $255\pi$ (medium and dark curves in Fig.~\ref{cavitypop}(b)), multiple peaks near resonance are seen due to the same effect and the increased skewing of the Rabi rotations from the stronger cavity drive. While these Rabi pulse areas are larger, we once again remark that the Rabi field will be much higher with a cavity drive for the same laser field, and experimentally one can plot the results as a function of peak laser power.

The same degree of exciton inversion can be achieved with cavity driving as with exciton driving by selecting the cavity drive strength. However, for a given drive, the scheme is more sensitive to changes in laser detuning with a higher-$Q$ cavity than with a lower-$Q$ cavity or no cavity. The cavity drive strength needed to achieve maximum inversion is higher for a lower-$Q$ cavity (scaling roughly with $1/Q$ as seen in these two examples). For a cavity drive strength $\Theta_c$, is as easy to produce as an exciton drive $\Theta_c/Q$, then a cavity with $Q = 9000$ or even $2000$ provides a significant enhancement compared to exciton driving alone.

Here the cavity was tuned $\Delta_{cx} = 0.83~\un{meV}$ above the exciton resonance, somewhat arbitrarily. If it is tuned instead to the frequency that produces the highest degree of inversion from exciton driving, $0.48~\un{meV}$ with these parameters, then that same degree of inversion can be reached with the lowest possible cavity drive strength. However, although there is an ideal cavity\textendash{}exciton detuning, the results are not very sensitive to small deviations from it, especially with lower-$Q$ cavities.

\section{Study and computed figures of merit for a phonon-assisted deterministic single photon source versus resonant $\pi$-pulse excitation }
Pulse-loaded exciton inversion in QDs may show promise as an implementation of a deterministic single photon source, and the question arises whether the phonon-assisted exciton inversion method would be useful in this regard. One key figure of merit is the indistinguishability of a single photon source, which can be quantified by the photon antibunching, the degree to which single photons are produced at regular time intervals rather than together. This can be calculated from the second-order two-time correlation function of a Hong--Ou--Mandel interferometer setup~\cite{PhysRevA.69.032305,He2013,PhysRevLett.59.2044}. The correlation function is defined formally as $G^{(2)}(t,\tau) = \braket{\sigp(t)\sigm(t+\tau)\sigp(t+\tau)\sigm(t)}$ and approximated as follows:
\begin{align}
G^{(2)}(t,\tau) &= \nonumber \\
&\frac{1}{2}\left(\braket{\sigp\sigm(t)}\braket{\sigp\sigm(t+\tau)} - |G^{(1)}(t,\tau)|^2\right),
\end{align}
where the first-order two-time correlation function is $G^{(1)}(t,\tau) = \braket{\sigp(t)\sigm(t+\tau)}$. The time-integrated $G^{(2)}$ function is $G^{(2)}(\tau) = \int_{-T}^{T}G^{(2)}(t,\tau)dt$, where $T$ is much longer than the pulse width and the radiative decay time.

To assess the single photon source figures of merit, we consider a third excitation scenario, distinct from the exciton- and cavity-driven systems considered previously. Here the QD is now exciton driven and its output is coupled to a cavity that is tuned on resonance ($\Delta_{cx} = 0$). This is similar to the scheme used in recent QD--cavity single photon source experiments~\cite{He2013,1367-2630-6-1-089}. It can be shown that, in the bad-cavity limit, the cavity and exciton populations are proportional: $a \propto \sigm, a^\dagger a \propto \sigp\sigm$, so that two-time correlations in the cavity photon number can be calculated based on the exciton number as described above.

The influence of the cavity is then incorporated as an enhanced spontaneous decay from the exciton to the cavity. The decay rate into the cavity is modified compared to the original spontaneous decay rate by the Purcell factor, $\tilde{\gamma} = F_P\gamma$, and the total decay from the exciton is $\gamma + \tilde{\gamma}$, which simply replaces $\gamma$ in the ME.
The on-resonance Purcell factor $F_P$ can be calculated as $2g^2/\gamma\kappa$, but it is treated here as an independent parameter to vary, since $g$ and $\kappa$ are not directly used in the bad cavity limit ME.

Based on the population results for the exciton-only system (Fig.~\ref{expth}(d)), we examine two-time correlations for two excitation scenarios: ($i$) resonant $\pi$-pulse excitation ($\Theta = \pi, \Delta_{Lx} = 0$), using the single Rabi flop of the coherent pulse to produce inversion; and ($ii$) off-resonant strong-pulse excitation ($\Theta = 18.7\pi, \Delta_{Lx} = 0.48~\un{meV}$), using the phonon bath to incoherently produce the QD inversion. Scenario ($ii$) is the mechanism studied throughout this work, but it is useful to compare it to the known mechanism of a standard $\pi$-pulse excitation. One advantage of scenario ($ii$) is that the single photons have a different frequency from the pump laser, allowing them to be spectrally isolated from the pump, which is ultimately required for an efficient single photon source from a solid state medium.

\begin{figure}[t]
\includegraphics[width=\linewidth]{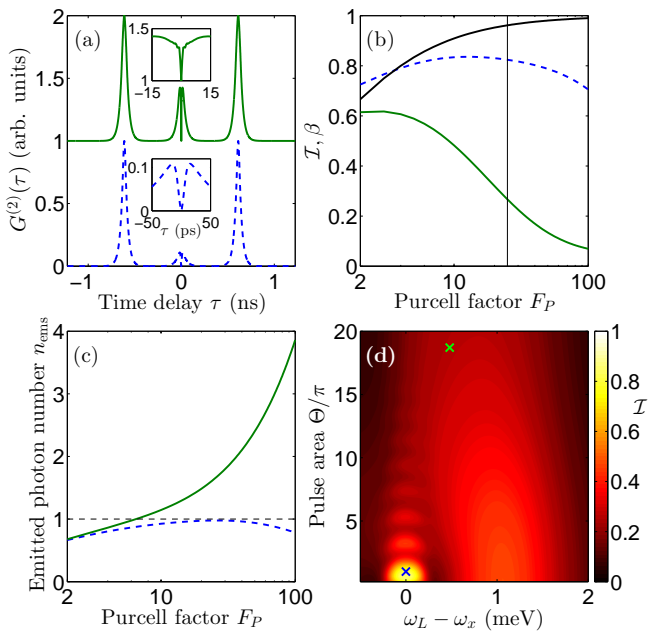}
\caption{\footnotesize{(Color online) (a) Second-order two-time correlation function $G^{(2)}(\tau)$ in the regime of $\pi$-pulse excitation ($\Theta = \pi, \Delta_{Lx} = 0$, blue dashed) and phonon-induced inversion ($\Theta = 18.7\pi, \Delta_{Lx} = 0.48~\un{meV}$, green solid), with $F_P = 10$. The insets show the behavior near $\tau = 0$, with $G^{(2)}(0) = 0$ by definition. (b) Photon antibunching, a measure of single photon indistinguishability, in the same two cases (blue dashed and green solid). Quantum efficiency $\beta$ (black) for varying $F_P$. The Purcell factor used in (d) is marked in thin black. (c) Emitted photon number in the two cases, which can be greater than 1 for off-resonant strong driving (green solid) but reaches a maximum of 0.977 with a Purcell factor of about 26 for $\pi$-pulse driving (blue dashed). (d) Photon antibunching with a Purcell factor of 25 for varying drive strengths and detunings, with the two cases considered in (a)-(c) marked in blue and green.}}
\label{G2}
\end{figure}

Figure~\ref{G2}(a) shows $G^{(2)}(\tau)$ for a train of pulses separated by $2T = 612~\un{ps}$, each with a pulse width of $18~\un{ps}$ and with Purcell factor $F_P = 10$ and other parameters as in Sec.~\ref{results_rates}. The correlation function is only computed for $\tau \ge 0$, but it is formally symmetric about $\tau = 0$ so the values are plotted for both positive and negative time delay for clarity. In Figure~\ref{G2}(b) we plot the indistinguishability, calculated as~\cite{PhysRevA.69.032305}
\begin{equation}
\mathcal{I} = 1 - \frac{\int_{-T}^{T}G^{(2)}(\tau)d\tau}{\int_{T}^{3T}G^{(2)}(\tau)d\tau}.
\end{equation}
That is, the indistinguishability is one minus the ratio of the area of the center peak in $G^{(2)}$ to the area of one of the side peaks. Also plotted in Fig.~\ref{G2}(b) is the collection efficiency $\beta = \frac{F_P}{1+F_P}$.

Another figure of merit of a single photon source is the number of photons emitted into the cavity with each input pulse~\cite{PhysRevA.69.032305}. Using cavity operators this would be defined as $n_{\rm ems} = \int_0^\infty\kappa\braket{a^\dagger a(t)}dt$, but using the proportionality of cavity and exciton operators it is given by $n_{\rm ems} = \int_0^\infty\tilde{\gamma}\braket{\sigp\sigm(t)}dt$. This is plotted in Fig.~\ref{G2}(c) for the two excitation cases. Finally, Fig.~\ref{G2}(d) plots the indistinguishability $\mathcal{I}$ for a Purcell factor of 25 in a range of excitation parameters.

In the regime of phonon-assisted population inversion, the indistinguishability is worse (as may be expected with a higher pump field) than in the regime of on-resonant $\pi$-pulse inversion, especially with higher Purcell factors. As well, a possible limitation exists that has been seen before for $\pi$-pulse excitation~\cite{PhysRevA.69.032305}: namely, it is difficult  to simultaneously achieve a very high quantum efficiency and a very high indistinguishability; one figure of merit is improved with high Purcell factors while the other relies on lower Purcell factors.
However, as mentioned in the introduction, 
recently  Somaschi {\em et al}.~\cite{Somaschi2015} has used
resonant $\pi$ pulse excitation of QDs in  micropillar cavities 
to achieve indistinguishabilities of close to unity (0.996). This increase
in comparison to our simulations may be caused by a lower
residual pure dephasing of the ZPL and possibly also due to 
the use of a narrowband spectral filter;  further work is needed
to explore and optimise the $\pi-$pulse excitation scheme which also uses a superposition of two orthogonal polarized excitons, but the experiments
are very encouraging.

Ideally, for a determinisctic excitation scheme,
the emitted photon number is expected to be as close as possible to 1 to achieve a single photon source. For the $\pi$-pulse excitation, this occurs with a Purcell factor of 26, where $n_{\rm ems} = 0.977$. In the phonon-assisted excitation regime, however, the emitted photon number exceeds 1 for Purcell factors above about 6. Indeed, it is entirely possible to emit multiple photons into the cavity if the exciton\textendash{}cavity decay rate $\tilde{\gamma}$ and the laser drive $\Theta$ are both strong enough: after the initial excitation of the dot, the first photon is emitted into the cavity while the laser pulse is still active, then the dot is excited again and emits a second photon into the cavity, and so on, creating up to four cavity photons when the Purcell factor is 100.\\


\section{Conclusions}
We have presented a  semi-analytical extension to polaron ME descriptions of phonon-assisted inversion in quantum-dot systems that is computationally simple, intuitively meaningful, and in good agreement with recent experiments showing the feasibility of such methods. The model was also extended and applied to describe a pulse-excited QD\textendash{}cavity system, which would provide either an enhancement of the excitation through cavity driving or a channel for photon collection for use as a deterministic single photon source.
As an application of the  theory, we also assessed the phonon-assisted loading scheme for use as a single photon source and generally conclude that  coherent loading schemes are likely better, such as those that use
Raman assisted processes~\cite{PhysRevB.82.045308}, adiabatic rapid passage~\cite{WeiNanoLett2014},  or
through direct $\pi$-pulse excitation with a spectral filter~\cite{Somaschi2015}.

\acknowledgments

This work was supported by the Natural Sciences and  Engineering Research Council of Canada and Queen's University. We thank Howard Carmichael and Mark Fox for useful discussions and John Quilter, Alistair Brash, and Feng Liu and Mark Fox from U. Sheffield for kindly providing the data shown in our Fig.~\ref{expth}.


\appendix*

\begin{widetext}
\section{Polaron master equation derivation}


%

%
%
%
%

The full (i.e., with no weak field approximation on the pump field) polaron ME for a QD\textendash{}phonon system is given by~\cite{PhysRevB.85.115309, McCutcheon2010}

\begin{equation}
\frac{\partial\rho}{\partial t} = \frac{1}{i\hbar}[H_S^\prime(t), \rho(t)] + \frac{\gamma}{2}\mathcal{L}[\sigm]\rho + \frac{\gamma^\prime}{2}\mathcal{L}[\sigp\sigm]\rho + \mathcal{L}_\mathrm{ph}\rho,
\end{equation}
with
\begin{equation}
\label{appFPME}
\mathcal{L}_\mathrm{ph}\rho =
 -\frac{1}{\hbar^2}\int_0^\infty\sum_{m=g,u}d\tau
 \left(G_m(\tau)[X_m(t), X_m(t,\tau)\rho(t)] +
 \operatorname{H.c.}\right),
\end{equation}
where $X_m(t,\tau) = e^{-iH_S^\prime(t)\tau/\hbar}X_m(t)e^{iH_S^\prime(t)\tau/\hbar}$. The phonon-modified system operators are
\begin{equation}
X_g(t) = \frac{1}{2}\hbar\Omega(t)(\sigp+\sigm) \qquad\text{and}\qquad
X_u(t) = \frac{1}{2}i\hbar\Omega(t)(\sigp-\sigm),
\end{equation}
and the polaron Green functions are
\begin{equation}
G_g(\tau) = \braket{B}^2\left(\cosh(\phi(\tau))-1\right) \qquad\text{and}\qquad
G_u(\tau) = \braket{B}^2\sinh(\phi(\tau)).
\end{equation}
Using the full polaron-transformed system Hamiltonian $H_S^\prime = -\hbar\Delta_{Lx}\sigp\sigm + \braket{B} X_g(t)$, we expand the two-time phonon system operators in terms of the one-time operators in the interaction picture.

We begin by noting that
\begin{align}
X_g(t) &= \frac{1}{2}\hbar\Omega(t)(\sigp+\sigm) \\
&= \frac{1}{2}\hbar\Omega(t)\sigma_x \\
X_g(t,\tau) &= \frac{1}{2}\hbar\Omega(t) e^{-iH_S^\prime(t)\tau/\hbar}\sigma_xe^{iH_S^\prime(t)\tau/\hbar},
\end{align}
and
\begin{align}
X_u(t) &= \frac{1}{2}i\hbar\Omega(t)(\sigp-\sigm) \\
&= -\frac{1}{2}\hbar\Omega(t)\sigma_y \\
X_u(t,\tau) &= -\frac{1}{2}\hbar\Omega(t) e^{-iH_S^\prime(t)\tau/\hbar}\sigma_ye^{iH_S^\prime(t)\tau/\hbar}.
\end{align}
We may then write~\cite{McCutcheon2010}
\begin{equation}
X_g(t,\tau) = \frac{1}{2}\hbar\Omega(t) \left(
 \frac{\Delta_{Lx}^2\cos(\eta(t)\tau) + \Omega_R(t)^2}{\eta(t)^2} \sigma_x
 - \frac{\Delta_{Lx}\sin(\eta(t)\tau)}{\eta(t)} \sigma_y
 - \frac{2\Delta_{Lx}\Omega_R(t)(1 - \cos(\eta(t)\tau))}{\eta(t)^2} \sigp\sigm
\right),
\end{equation}
and
\begin{equation}
X_u(t,\tau) = -\frac{1}{2}\hbar\Omega(t) \left(
 \frac{\Delta_{Lx}\sin(\eta(t)\tau)}{\eta(t)} \sigma_x
 + \cos(\eta(t)\tau) \sigma_y
 + \frac{2\Omega_R(t)\sin(\eta(t)\tau)}{\eta(t)} \sigp\sigm
\right),
\end{equation}
where $\eta(t) = \sqrt{\Omega_R(t)^2 + \Delta_{Lx}^2}$ and $\Omega_R(t) = \braket{B}\Omega(t)$.
These expanded analytical expressions for the system operators are then used to separate the commutators in Eq.~\ref{appFPME} to produce Lindblad and non-Lindblad terms of $\sigp$, $\sigm$ and $\rho$. We proceed as follows:
\begin{align}
\mathcal{L}_\mathrm{ph}\rho 
 &=
  -\frac{1}{\hbar^2}\int_0^\infty
  \left(G_g(\tau)\left[X_g(t), X_g(t,\tau)\rho(t)\right] +
  \operatorname{H.c.} +
  G_u(\tau)\left[X_u(t), X_u(t,\tau)\rho(t)\right] +
  \operatorname{H.c.}\right)
  d\tau, \\
\
 \begin{split}
 &=
  -\frac{1}{\hbar^2}\int_0^\infty \Bigg(
  G_g(\tau)\bigg[\frac{1}{2}\hbar\Omega(t)(\sigp+\sigm),
  \frac{1}{2}\hbar\Omega(t) \bigg(
   \frac{\Delta_{Lx}^2\cos(\eta(t)\tau) + \Omega_R(t)^2}{\eta(t)^2} \sigma_x \\
   &\hspace{104pt}- \frac{\Delta_{Lx}\sin(\eta(t)\tau)}{\eta(t)} \sigma_y
   - \frac{2\Delta_{Lx}\Omega_R(t)(1 - \cos(\eta(t)\tau))}{\eta(t)^2} \sigp\sigm
  \bigg)\rho(t)\bigg] +
  \operatorname{H.c.} \\
  &\hspace{52pt}+
  G_u(\tau)\bigg[\frac{1}{2}i\hbar\Omega(t)(\sigp-\sigm),
  -\frac{1}{2}\hbar\Omega(t) \bigg(
   \frac{\Delta_{Lx}\sin(\eta(t)\tau)}{\eta(t)} \sigma_x \\
   &\hspace{104pt}+ \cos(\eta(t)\tau) \sigma_y
   + \frac{2\Omega_R(t)\sin(\eta(t)\tau)}{\eta(t)} \sigp\sigm
  \bigg)\rho(t)\bigg] +
  \operatorname{H.c.}
  \Bigg) d\tau.
 \end{split}
\end{align}
For convenience we define shorthand formulas for expressions that appear frequently in these equations, all of which are real, dimensionless, and scalar functions of $t$ and $\tau$:
\begin{align}
f(t,\tau) &= \frac{\Delta_{Lx}^2\cos(\eta(t)\tau) + \Omega_R(t)^2}{\eta(t)^2}, \\
g(t,\tau) &= \frac{\Delta_{Lx}\sin(\eta(t)\tau)}{\eta(t)}, \\
h(t,\tau) &= \frac{2\Delta_{Lx}\Omega_R(t)(1 - \cos(\eta(t)\tau))}{\eta(t)^2}, \\
q(t,\tau) &= \cos(\eta(t)\tau), \\
r(t,\tau) &= \frac{2\Omega_R(t)\sin(\eta(t)\tau)}{\eta(t)}.
\end{align}
Consequently we can write
\begin{align}
 \begin{split}
 \mathcal{L}_\mathrm{ph}\rho
 &=
  -\frac{\Omega(t)^2}{4}\int_0^\infty \Bigg(
  G_g(\tau)\bigg[(\sigp+\sigm),
  \bigg(
   f(t,\tau) (\sigp+\sigm)
   + ig(t,\tau) (\sigp-\sigm)
   - h(t,\tau) \sigp\sigm
  \bigg)\rho(t)\bigg] +
  \operatorname{H.c.} \\
  &\hspace{65pt}+
  G_u(\tau)\bigg[i(\sigp-\sigm),
  -\bigg(
   g(t,\tau) (\sigp+\sigm)
   - iq(t,\tau) (\sigp-\sigm)
   + r(t,\tau) \sigp\sigm
  \bigg)\rho(t)\bigg] +
  \operatorname{H.c.}
  \Bigg) d\tau,
 \end{split} \\
\
 \begin{split}
 &=
 -\frac{\Omega(t)^2}{4}\int_0^\infty \Bigg(
  G_g(\tau)\bigg(
   f(t,\tau)
    [(\sigp+\sigm), (\sigp+\sigm)\rho(t)]
   +g(t,\tau)
    [(\sigp+\sigm), i(\sigp-\sigm)\rho(t)] \\
   &\hspace{98pt}-h(t,\tau)
    [(\sigp+\sigm), \sigp\sigm\rho(t)]
  \bigg) + \operatorname{H.c.} \\
  &\hspace{52pt}+
  G_u(\tau)\bigg(
  -g(t,\tau)
   [i(\sigp-\sigm), (\sigp+\sigm)\rho(t)]
  +q(t,\tau)
   [i(\sigp-\sigm), i(\sigp-\sigm)\rho(t)] \\
  &\hspace{98pt}-r(t,\tau)
   [i(\sigp-\sigm), \sigp\sigm\rho(t)]
  \bigg) + \operatorname{H.c.}
 \Bigg) d\tau,
 \end{split}
\end{align}
\begin{align}
 \begin{split}
 &=
 -\frac{\Omega(t)^2}{4}\int_0^\infty \bigg(
  -\operatorname{Re}\{G_g(\tau)\}
   f(t,\tau) \left(
    \mathcal{L}[\sigp]\rho
    +\mathcal{L}[\sigm]\rho
    +2(\sigp\rho\sigp + \sigm\rho\sigm)
    \right) \\
   &\hspace{71pt}-i\operatorname{Re}\{G_g(\tau)\}
   g(t,\tau) \left(
     2(\sigp\rho\sigp - \sigm\rho\sigm)
    +2[\sigp\sigm,\rho]
    \right) \\
   &\hspace{71pt}+~\operatorname{Im}\{G_g(\tau)\}
   g(t,\tau) \left(
     \mathcal{L}[\sigp]\rho
    -\mathcal{L}[\sigm]\rho
    \right) \\
   &\hspace{71pt}-~\operatorname{Re}\{G_g(\tau)\}
   h(t,\tau) \left(
    -\sigp\sigm\rho\sigp - \sigm\rho\sigp\sigm
    +\sigm\rho - \sigp\sigm\rho\sigm + \rho\sigp - \sigp\rho\sigp\sigm
    \right) \\
   &\hspace{71pt}-i\operatorname{Im}\{G_g(\tau)\}
   h(t,\tau) \left(
    -\sigp\sigm\rho\sigp + \sigm\rho\sigp\sigm
    +\sigm\rho - \sigp\sigm\rho\sigm - \rho\sigp + \sigp\rho\sigp\sigm
    \right) \\
  &\hspace{71pt}
   +i\operatorname{Re}\{G_u(\tau)\}
   g(t,\tau) \left(
    2(\sigp\rho\sigp - \sigm\rho\sigm)
   -2[\sigp\sigm,\rho]
    \right) \\
   &\hspace{71pt}+~\operatorname{Im}\{G_u(\tau)\}
   g(t,\tau) \left(
     \mathcal{L}[\sigp]\rho
    -\mathcal{L}[\sigm]\rho
    \right) \\
   &\hspace{71pt}-~\operatorname{Re}\{G_u(\tau)\}
   q(t,\tau) \left(
     \mathcal{L}[\sigp]\rho
    +\mathcal{L}[\sigm]\rho
    -2(\sigm\rho\sigm + \sigp\rho\sigp)
    \right) \\
  &\hspace{71pt}-i\operatorname{Re}\{G_u(\tau)\}
  r(t,\tau) \left(
    -\sigp\sigm\rho\sigp + \sigm\rho\sigp\sigm
    -\sigm\rho + \sigp\sigm\rho\sigm + \rho\sigp - \sigp\rho\sigp\sigm
    \right) \\
  &\hspace{71pt}+~\operatorname{Im}\{G_u(\tau)\}
  r(t,\tau) \left(
    -\sigp\sigm\rho\sigp - \sigm\rho\sigp\sigm
    -\sigm\rho + \sigp\sigm\rho\sigm - \rho\sigp + \sigp\rho\sigp\sigm
    \right)
 \bigg) d\tau.
 \end{split}
\end{align}

Noting that the operators are independent of $\tau$, we isolate the phonon rates:
\begin{align}
\begin{split}
\mathcal{L}_\mathrm{ph}\rho
 &=
    \frac{\Gamma^{\sigma^+}}{2}\mathcal{L}[\sigp]\rho 
  + \frac{\Gamma^{\sigma^-}}{2}\mathcal{L}[\sigm]\rho
  - \Gamma^\mathrm{cd}(\sigp\rho\sigp + \sigm\rho\sigm)
  -i\Gamma^\mathrm{sd}(\sigp\rho\sigp - \sigm\rho\sigm)
  +i\Delta^{\sigp\sigm} [\sigp\sigm,\rho] \\
 &~~~~- \left(i \Gamma_u(\sigp\sigm\rho\sigp + \sigm\rho - \sigp\sigm\rho\sigm) + \operatorname{H.c.}\right)
  - \left(\Gamma_g(\sigp\sigm\rho\sigp - \sigm\rho + \sigp\sigm\rho\sigm) + \operatorname{H.c.}\right),
\end{split}
\label{appAPME}
\end{align}
where
\begin{align}
\Gamma^{\sigma^+}
 &=
  -\frac{\Omega(t)^2}{2}\int_0^\infty \bigg(
   - \operatorname{Re}\{G_g(\tau)\}f(t,\tau)
   + \operatorname{Im}\{G_g(\tau) + G_u(\tau)\}g(t,\tau)
   - \operatorname{Re}\{G_u(\tau)\}q(t,\tau)
  \bigg) d\tau, \\
\
 \begin{split}
 &=
  -\frac{\Omega(t)^2}{2}\int_0^\infty \bigg(
   - \operatorname{Re}\{\braket{B}^2\left(\cosh(\phi(\tau))-1\right)\}
     \left(\frac{\Delta_{Lx}^2\cos(\eta(t)\tau) + \Omega_R(t)^2}{\eta(t)^2} \right) \\
   &\hspace{76pt}+ \operatorname{Im}\{\braket{B}^2\left(\cosh(\phi(\tau))-1\right) + \braket{B}^2\sinh(\phi(\tau))\}
     \left(\frac{\Delta_{Lx}\sin(\eta(t)\tau)}{\eta(t)}\right) \\
   &\hspace{76pt}- \operatorname{Re}\{\braket{B}^2\sinh(\phi(\tau))\}
     \left(\cos(\eta(t)\tau)\right)
  \bigg) d\tau,
 \end{split} \\
\
 \begin{split}
 &=
  \frac{\Omega_R(t)^2}{2}\int_0^\infty \bigg(
   \operatorname{Re}\left\{\left(\cosh(\phi(\tau))-1\right)
     \left(\frac{\Delta_{Lx}^2\cos(\eta(t)\tau) + \Omega_R(t)^2}{\eta(t)^2} \right) 
    + \sinh(\phi(\tau)) \left(\cos(\eta(t)\tau)\right) \right\} \\
   &\hspace{76pt}- \operatorname{Im}\{e^{\phi(\tau)}-1\}
     \left(\frac{\Delta_{Lx}\sin(\eta(t)\tau)}{\eta(t)}\right)
  \bigg) d\tau.
 \end{split}
\end{align}
Similarly for the other rates,
\begin{align}
\Gamma^{\sigma^-}
 &=
  \frac{\Omega_R(t)^2}{2}\int_0^\infty \bigg(
   \operatorname{Re}\left\{\left(\cosh(\phi(\tau))-1\right)
     \left(\frac{\Delta_{Lx}^2\cos(\eta(t)\tau) + \Omega_R(t)^2}{\eta(t)^2} \right) 
    + \sinh(\phi(\tau)) \left(\cos(\eta(t)\tau)\right) \right\} \nonumber \\
   &\hspace{76pt}+ \operatorname{Im}\{e^{\phi(\tau)}-1\}
     \left(\frac{\Delta_{Lx}\sin(\eta(t)\tau)}{\eta(t)}\right)
  \bigg) d\tau,
 \end{align}
%
\begin{align}
 \begin{split}
 \Gamma^{\mathrm{cd}}
 &=
  \frac{\Omega_R(t)^2}{2}\int_0^\infty
   \operatorname{Re}\left\{
   \sinh(\phi(\tau)) \left(\cos(\eta(t)\tau)\right)
  -\left(\cosh(\phi(\tau))-1\right)
    \left(\frac{\Delta_{Lx}^2\cos(\eta(t)\tau) + \Omega_R(t)^2}{\eta(t)^2}\right)
   \right\}
  d\tau,
 \end{split}
\end{align}
\begin{align}
\Gamma^{\mathrm{sd}}
 \begin{split}
 &=
  \frac{\Omega_R(t)^2}{2}\int_0^\infty \bigg(
    \operatorname{Re}\{1 - e^{-\phi(\tau)}\}
    \left( \frac{\Delta_{Lx}\sin(\eta(t)\tau)}{\eta(t)} \right)
  \bigg) d\tau,
 \end{split}
\end{align}
\begin{align}
\Delta^{\sigp\sigm}
 \begin{split}
 &=
  \frac{\Omega_R(t)^2}{2}\int_0^\infty
    \operatorname{Re}\{e^{\phi(\tau)}-1\}
    \left( \frac{\Delta_{Lx}\sin(\eta(t)\tau)}{\eta(t)} \right)
  d\tau,
 \end{split}
\end{align}
\begin{align}
\Gamma_g
 &=
  \frac{\Omega_R(t)^2}{2}\int_0^\infty
  \left(\cosh(\phi(\tau))-1\right)
  \frac{\Delta_{Lx}\Omega_R(t)(1 - \cos(\eta(t)\tau))}{\eta(t)^2}
  d\tau,
\Gamma_u
 &=
  \frac{\Omega_R(t)^2}{2}\int_0^\infty
  \sinh(\phi(\tau))
  \frac{\Omega_R(t)\sin(\eta(t)\tau)}{\eta(t)}
  d\tau.
\end{align}

These phonon-mediated scattering rates along with Eq.~\ref{appAPME} form the semi-analytical polaron ME used in the main text. Each analytical rate is a function of $t$, and its value at each time $t$ is calculated by integrating with respect to $\tau$; however, this is much less computationally expensive than the operator exponentials inside the time integral in Eq.~\ref{appFPME}. Thus, this analytical restructuring of the ME, while somewhat complicated, reduces greatly the time required to compute the phonon contribution. It also separates the phonon contribution into meaningful rates corresponding to phonon emission, phonon absorption, dephasing, frequency shift, and other processes that help to explain the underlying physics of electron-phonon coupling in these optically driven QD systems.\\


\end{widetext}

\bibliographystyle{unsrt}
\bibliography{refs}

\end{document}